\documentclass{jors}
\usepackage{natbib}
\usepackage{amsmath,amssymb,amsfonts}
\usepackage{listings}
\usepackage{algorithmic}
\usepackage{bm}
\usepackage{graphicx}
\usepackage{textcomp}
\usepackage{subcaption}
\usepackage{siunitx}
\usepackage{float}
\usepackage{svg}
\usepackage{caption}
\usepackage{makecell}
\newcommand{\rev}[1]{#1}
\definecolor{Red}{rgb}{1,0,0}
\definecolor{codegreen}{rgb}{0,0.6,0}
\definecolor{codegray}{rgb}{0.5,0.5,0.5}
\definecolor{codepurple}{rgb}{0.58,0,0.82}
\definecolor{backcolour}{rgb}{0.95,0.95,0.92}
\lstdefinestyle{mystyle}{
  backgroundcolor=\color{backcolour},
  commentstyle=\color{codegreen},
  keywordstyle=\color{magenta},
  numberstyle=\tiny\color{codegray},
  stringstyle=\color{codepurple},
  basicstyle=\footnotesize,
  breakatwhitespace=false,
  breaklines=true,
  captionpos=b,
  keepspaces=false,
  numbers=left,
  numbersep=1pt,
  showspaces=false,
  showstringspaces=false,
  showtabs=false,
  tabsize=2,
  columns=fullflexible
}
\definecolor{mygray}{gray}{0.6}

\begin{document}

{\bf Integrating Odeint Time Stepping into OpenFPM for Distributed and GPU Accelerated Numerical Solvers} \\

\rule{\textwidth}{1pt}

\section*{Paper Authors}

Abhinav Singh$^{1,2,3,4}$ (Lead author) \\ Landfried Kraatz$^1$ \\ Serhii Yaskovets$^{1,2,3}$ \\ Pietro Incardona$^{1,2,3}$ \\ Ivo F.~Sbalzarini$^{1,2,3,4}$ (Corresponding author, \url{sbalzarini@mpi-cbg.de})

\section*{Paper Author Roles and Affiliations}
\begin{enumerate}
  \item Dresden University of Technology, Faculty of Computer Science, Dresden, Germany.
  \item Max Planck Institute of Molecular Cell Biology and Genetics, Dresden, Germany.
  \item Center for Systems Biology Dresden, Dresden, Germany.
  \item Center for Scalable Data Analytics and Artificial Intelligence (ScaDS.AI), Dresden/Leipzig.
\end{enumerate}

\section*{Abstract}
We present a software implementation integrating the time-integration library Odeint from Boost with the OpenFPM framework for scalable scientific computing. This enables compact and scalable codes for multi-stage, multi-step, and adaptive explicit time integration on distributed-memory parallel computers and on (clusters of) Graphics Processing Units (GPUs). The present implementation is based on extending OpenFPM's metaprogramming system to Odeint data types. This makes the time-integration methods from Odeint available in a concise template-expression language for numerical simulations distributed and parallelized using OpenFPM. We benchmark the present software for exponential and sigmoidal dynamics and present an application example to the 3D Gray-Scott reaction-diffusion problem portable to both CPUs and GPUs using only 60 lines of code.

\section*{Keywords}
numerical integration; scientific computing; parallel computing; differential equations; GPU acceleration; distributed computing; numerical methods

\section*{(1) Overview}

\vspace{0.5cm}

\section*{Introduction}

High-performance computers are commonly used for computer simulations of dynamical
models and for numerically solving mathematical equations describing the dynamics of
a system or process. This is done using time-integration methods, which are numerical
algorithms that approximate the trajectory of the state of a dynamical system to
a future point in time. Model dynamics is simulated by iteratively taking
(sufficiently small) time steps. Methods that, in each iteration, extrapolate the
next time step from the current and past ones are called {\em explicit} time
integrators, or {\em explicit} time-stepping schemes. They successively advance the state
$\mathbf{u}(t)$ of the simulation from a time $t$ to a time $t+\delta t$, $\delta t > 0$, given an expression for its time derivative
\[
  \frac{\mathrm{d}\mathbf{u}}{\mathrm{d}t} = \mathcal{F}(t,\mathbf{u}(t))
\]
and starting from an initial condition $\mathbf{u}(0) = \mathbf{u}_0$. The right-hand
side $\mathcal{F}$ is given by the model that is solved or simulated, or it results
from spatial discretization of a partial differential equation (PDE). In order for
the thus-computed sequence $\hat{\mathbf{u}}_0 = \mathbf{u}(0\delta
t),\,\hat{\mathbf{u}}_1
\approx \mathbf{u}(1\delta t),\,\hat{\mathbf{u}}_2 \approx \mathbf{u}(2\delta t),\,\ldots$ to converge to the true dynamics $\mathbf{u}(t)$ when $\delta t \to 0$, a time-integration method must be both {\em consistent} and {\em stable}.

In simulations of nonlinear dynamics, identifying a consistent and stable
time-integration method often requires experimenting with different step sizes and
stepping algorithms, or autotuning the choices \citep{khouzami_jocs21}.
In a distributed-memory codes, the state $\mathbf{u}$ needs to be communicated between processes at inter-process boundaries.
Since the time-step size is limited by the fastest dynamics to be resolved, methods with uniform step sizes are often wasteful, and instead adaptive methods are used \citep{press1992adaptive}, which adjust the step size to the simulated dynamics.
This requires additional global communication of error estimates and step sizes among all processes.
Both communications, state boundaries and global step size, need to be implemented, usually leading to method-specific code.

Here, we reduce the need for writing method-specific code by presenting a software design and implementation for integrating the time-stepping library Boost Odeint~\citep{Ahnert2011,odeint2015ahnert} with the scalable numerical simulation library OpenFPM~\citep{INCARDONA2019155}. OpenFPM is an open-source software library for implementing
scalable particle and hybrid particle--mesh simulations on heterogeneous
high-performance architectures including CPU clusters and (clusters of) GPUs. It uses C++ template metaprogramming to provide hardware-abstracted data structures and performance-portable algorithms~\citep{Incardona:2023a}.
In order to reduce development times for numerical solvers, OpenFPM provides
a template expression system to encode PDEs in near-mathematical
notation~\citep{singh_c_2021}. We complement this with a template system that allows
OpenFPM to interface with Odeint. This enables the use of abstract encapsulation of
time derivatives in a simulation model, which can directly be used by Odeint for time
stepping with a variety of high-order and adaptive time-integration methods.
Importantly, the present implementation follows the software design principle of
separation of concerns. This means that the implementation of a distributed
time-stepping scheme is not specific or limited to a given right-hand side
$\mathcal{F}$, nor are the spatial operators used to discretize a PDE within
$\mathcal{F}$ explicit in the time-integration code. This allows modular combinations
of right-hand sides and time-stepping methods, independent of the spatial
discretizaton and the computer architecture used.

\section*{Implementation and architecture}

We make the explicit time-stepping schemes from {\texttt{Boost::odeint}}~\citep{odeint2015ahnert} available in OpenFPM~\citep{INCARDONA2019155} by designing and implementing an interface between these two software libraries. This interface is based on extending Odeint to OpenFPM data types and making the Odeint concepts available in OpenFPM's template metaprogramming expression system. Both OpenFPM and Odeint are templated C++ libraries.

\subsection*{Architecture and nomenclature of Boost Odeint}

Boost Odeint~\citep{odeint2015ahnert} offers a selection of time-stepping methods to numerically solve initial-value problems, and it allows for
implementing custom time steppers. Odeint uses template metaprogramming (TMP) to
render the implementation independent of the underlying data structures. The key data
structure is the \emph{state} of the ODE system. It is of the abstract data type
\emph{state-type}, defined by a template parameter at compile time. The \emph{state}
represents the solution of the ODE \emph{System} at a given time. To use Odeint, an
initial \emph{state} has to be provided as an input, along with a \emph{System}
function or a TMP functor that computes the right-hand side $\mathcal{F}$ for the current \emph{state} $\mathbf{u}$. One can then use any available time stepper from Odeint, as summarized in
Table 1, to extrapolate the \emph{state} to the next time step and update it in-place.

When using adaptive time-stepping methods, the size of each time-step is determined from an error estimator and a predefined error tolerance. In Odeint, time steppers that can estimate the error in the solution are called \emph{error stepper}s. Error-controlled adaptive steppers can be constructed for any \emph{error stepper}~\citep{odeint2015ahnert}.

All time steps in Odeint are instantiated by a constructor specifying what Odeint calls an \emph{Algebra} for the \emph{state}. The constructor creates temporary variables of the same \emph{state-type} as required for a specific stepper. The Odeint \emph{Algebra} defines arithmetic operations on the \emph{state-type}, allowing Odeint to compute linear combinations of \emph{state}s to, e.g., combine the stages in a multi-stage time integrator.

\begin{figure}
  \centering
  \includegraphics[scale=0.4, trim=10pt 10pt 10pt 10pt,clip]{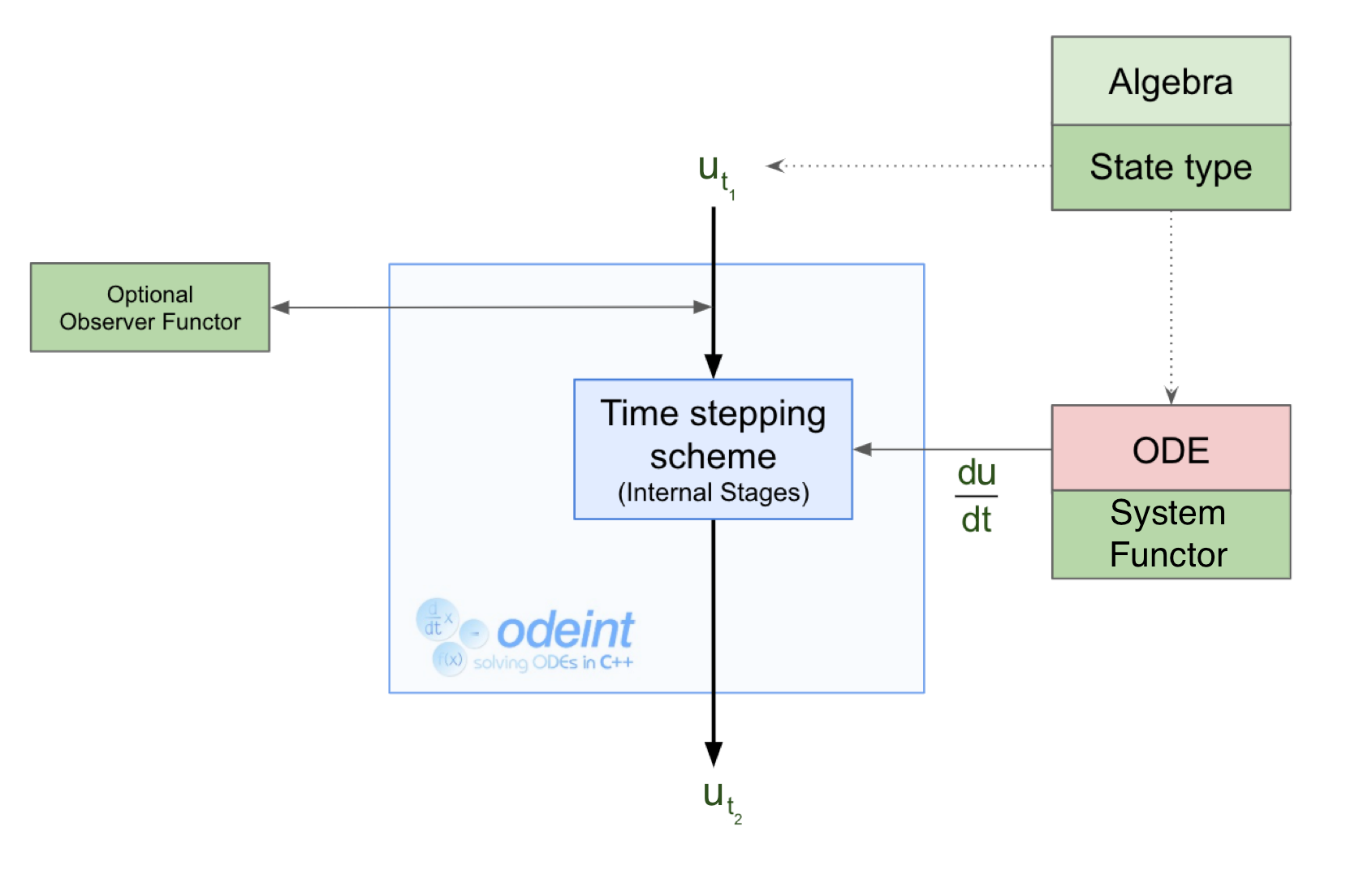}
  \caption{Architecture of the Odeint Library. In each iteration, the \emph{state}
    $\mathbf{u}$ is advanced from a time $t_1$ to a later time $t_2=t_1+\delta t$,
    $\delta t >0$. The right-hand side $\mathcal{F}(t,\mathbf{u}(t))$ of the ODE
    system is encapsulated in the \emph{System} functor. An \emph{Algebra} defines
    mathematical operations over the \emph{state-type}, i.e., the data type of the
    \emph{state}. An optional \emph{Observer} functor can execute user code at the
  beginning of every time step.}
  \label{fig:odeint}
\end{figure}

To use adaptive or error-controlled steppers, the \emph{state} object must also provide a method to compute a norm over the corresponding \emph{state-type}. This norm is usually the $L_\infty$ norm to bound the global error, but any other norm can alternatively be implemented. In addition, a separate function or functor can be passed to Odeint as an \emph{Observer}, which is called before each time step. This can be used, e.g., to write output to a file or to provide additional code to be run before each time step. The interplay of these objects in Odeint is illustrated in Fig.~\ref{fig:odeint}.

By default Odeint supports \emph{state-type}s such as
{\texttt{std::vector}} and {\texttt{boost::array}}. None of the default
\emph{state-type}s, however, allow for distributed computing on heterogeneous
architectures. However, user-defined custom \emph{state-type}s can be implemented.
Here, we leverage this facility to provide an OpenFPM \emph{state-type} for Odeint
that internally encapsulates all parallel and distributed computing
operations.

\begin{center}
  \captionof{table}{Explicit time-stepping schemes available in Odeint.\label{Tab:steppers}}
  \vspace{0.5em} 
  \begin{tabular}{|c|c|c|}
    \hline
    Category & Method & Accuracy order \\
    \hline\hline
    fixed step size & explicit Euler & 1 \\
    & modified midpoint & 2 \\
    & Runge--Kutta 4 & 4 \\
    & Runge--Kutta--Cash--Karp 54 & 5 \\
    & Runge--Kutta--Dopri 5 & 5 \\
    & Runge--Kutta--Fehlberg 78 & 8 \\
    \hline
    dynamic step size & Runge--Kutta--Cash--Karp 54 & 5 \\
    & Runge--Kutta--Dopri 5 & 5 \\
    & Runge--Kutta--Fehlberg 78 & 8 \\
    \hline
    multi-step & Adams--Bashforth & $1\ldots8$ \\
    & Adams--Bashforth--Moulton & $1\ldots8$ \\
    \hline
    symplectic & symplectic Euler & 1 \\
    & velocity Verlet & 2 \\
    & Runge--Kutta--McLachlan & 4 \\
    \hline
  \end{tabular}
\end{center}

\subsection*{The distributed OpenFPM state type and algebra}

We provide a custom Odeint \emph{state-type}, along with
its \emph{Algebra}, for the distributed data types of OpenFPM. This OpenFPM \emph{state-type} encapsulates OpenFPM's abstract data structures and performance-portable algorithms for shared- and distributed-memory computing, as well as multi-GPU computing. OpenFPM is implemented in C++ using \rev{template metaprogramming} \citep{Incardona:2023a}, making it compatible with Odeint in style and architecture. The custom Odeint
\emph{state-type}s for OpenFPM are internally distributed and specialized for different hardware backends. This includes Nvidia and AMD GPU backends~\citep{Incardona:2023a}.
Specifically, we provide Odeint \emph{state-type}s and \emph{Algebra}s for OpenFPM
distributed vectors on CPUs and GPUs. These \emph{state-type}s are named
{\texttt{state\textunderscore type\textunderscore\#d\textunderscore
ofp[\textunderscore gpu]}}, where {\texttt{\#}} is substituted with the natural
number specifying the dimensionality of the vectors in the array. The current implementation supports array dimensions between 1 and 6. For example, the
\emph{state-type} for a scalar field on the CPU is {\texttt{state\textunderscore
type\textunderscore1d\textunderscore ofp}} and for a 3D vector field on the GPU it is
{\texttt{state\textunderscore type\textunderscore3d\textunderscore ofp\textunderscore
gpu}}.

with a discretization that represents space (every discretization point is an element of the \emph{state-type})

\rev{By default, Odeint compiles any custom \emph{state-type} with a default \emph{Algebra}. This default \emph{Algebra}, however, does not consider the domain decomposition in a distributed computer, which requires synchronization before time-stepping. Since OpenFPM data types are internally and transparently distributed, a custom \emph{Algebra} is required. This custom \emph{Algebra} must implement the required arithmetic operations over internally distributed OpenFPM data containers. For this, every discretization element (grid point or particle) in an OpenFPM container becomes an element of the Odeint \emph{state-type}. The custom \emph{Algebra} then accounts for domain decomposition by partitioning the \emph{state-type} across MPI processes, transparently managing ghost (overlap, halo) zones for inter-process synchronization.
}

\rev{For GPU-enabled state types, we create a custom Odeint \emph{Operations} class. The encapsulated operations are dispatched to CUDA (for Nvidia GPUs) or HIP (for AMD GPUs) kernels, ensuring that element-wise computations and reductions (e.g., norm evaluations) are executed efficiently on the GPU. Our custom class ensures performance portability across heterogeneous architectures and provides a practical blueprint for developers looking to implement custom Odeint \emph{Algebra}s for distributed or multi-GPU environments.}

\begin{lstlisting}[caption={The OpenFPM distributed vector \emph{state-type} for Odeint.},label={lst:state_type},language=C++]
  //Templated with OpenFPM distributed vector type
  template<typename vector_type>
  struct state_type_2d_ofp{
    state_type(){}
    typedef size_t size_type;
    typedef int is_state_vector;
    aggregate<texp_v<double>,texp_v<double>> data;

    //Method to get the size
    size_t size() const
    { return data.get<0>().size(); }

    //Method to resize
    void resize(size_t n)
    {
      data.get<0>().resize(n);
      data.get<1>().resize(n);
    }
  };

  //Additional structs as required by Odeint
  namespace boost::numeric::odeint {
    template<>
    struct is_resizeable<state_type<vector_type>> {
      typedef boost::true_type type;
      static const bool value = type::value;
    };
    template<typename T>
    struct vector_space_norm_inf<state_type<T>>
    {
      typedef typename T::stype result_type;
    };
  }
\end{lstlisting}

Listing~\ref{lst:state_type} shows as an example the \emph{state-type} for an OpenFPM
distributed vector on the CPU with two components of type {\texttt{double}} expressed as an {\texttt{aggregate}} (Line 7).
The container {\texttt{texp\_v}} is a special container in OpenFPM's embedded
domain-specific language (DSL) for ODEs and PDEs~\citep{singh_c_2021}. It allows array and tensor operations to be expressed using MATLAB \citep{MATLAB} or Numpy \citep{Numpy} syntax \citep{singh_c_2021}. It also supports implicit mathematical equations that require matrix assembly and numerical solution of a system of algebraic equations. It is therefore a convenient choice for storing (temporary) results in a way that is compatible with the operators of the embedded DSL, except operations that require inter-process communication \citep{singh_c_2021}.
Because {\texttt{texp\_v}} is a resizable object, we notify Boost that the entire \emph{state-type} is resizable (Lines {22--25}). In addition, since adaptive steppers require calculating a norm of the \emph{state}, we specify the result type of the infinity norm (Line 28).

The custom \emph{Algebra} for this OpenFPM \emph{state-type} is called {\texttt{odeint::vector\_space\_algebra\_ofp}}. It is made available by the header file {\texttt{OdeIntegrators/}}-{\texttt{vector\_algebra\_ofp.hpp}} and defines the distributed mathematical operations on OpenFPM distributed vectors. It needs to be specified at compile time (cf.~Listing \ref{lst:time-stepper-example}, Line 6).

The OpenFPM distributed \emph{Algebra} for Odeint further defines the iterators {\texttt{for\_each\#()}} and {\texttt{for\_each\_prop\#()}} as inline methods using OpenFPM's distributed iterators. This encapsulates algebraic operations over discretization elements (i.e., grid points or particles) or their individual properties in multi-dimensional fields, respectively. \rev{In these iterators, {\texttt{\#}} is the order of the operation. For example, if the iterator is to apply a unary operator across all elements of a \emph{state-type}, {\texttt{\#=1}}. For a binary operator across pairs of state elements, {\texttt{\#=2}}, and so on. The time-integration methods of Odeint use these operator iterators. The most complex Odeint integrators require operators across up to {\texttt{\#=15}} state elements, e.g., for an eighth-order adaptive multi-stage scheme.}

Finally, the OpenFPM distributed \emph{algebra} for Odeint provides the methods \texttt{for\_each\_norm()} and \texttt{for\_each\_prop\_resize()} for
computing \emph{state} norms in adaptive time steppers and for distributed \emph{state}
object resizing, respectively. All right-hand-side-specific and method-specific code
is  generated by the compiler at compile time from this
\emph{Algebra} in conjunction with the OpenFPM template expression system for
differential equations~\citep{singh_c_2021}, ensuring modularity of the design and
performance portability over different hardware architectures. \rev{Together, these operations enable the efficient, inlined computations required for intermediate stage evaluations of explicit multi-stage schemes.}

\section*{Quality control}

We use a Continuous Integration / Continuous Delivery (CI/CD) development methodology in order to accelerate the software development life cycle and improve the overall quality of the code. In our CI/CD pipeline, we systematically assure compatibility with multiple compilers (GNU Compiler Collection, Clang Compiler, Intel Compiler), operating systems (Linux, UNIX-like, macOS, Windows/Cygwin), and hardware (x86\_64, AMD64, ARM64, Nvidia GPU). As an orchestration tool, we use Rundeck. It supports automatic overnight builds and deployment, has interfaces to Gitlab and Github to enable repository mirroring, and it has an interface to the official OpenFPM website \url{https://openfpm.mpi-cbg.de} to upload up-to-date reports on runtime performance, static analysis, and test coverage. The website also contains information about compiling the software from source, installing the software in a Docker container, and how to use the examples supplied with the package. In addition, Doxygen code documentation is available at \url{https://ppmcore.mpi-cbg.de/doxygen/openfpm/}. The documentation is automatically generated from the source code with every build. Since test coverage is difficult to evaluate for templated C++ code, we report the coverage as estimated by the tool {\tt gcov}. We ensure this coverage to be above 90\% for each OpenFPM submodule. The test suite consists of more than 600 unit tests covering more than 95\% of the entire OpenFPM project. Finally, our CI/CD pipeline facilitates code dissemination by providing automatically built Docker container images. This improves portability and lowers the entry barrier for new users.

Using the code thus made available, we next present accuracy benchmarks, confirming the mathematical correctness of our implementation, and of scalability benchmarks, testing the parallel efficiency of the implementation on CPUs and GPUs.

\subsection*{Accuracy Benchmarks}\label{sec:usecase1}

\begin{figure}[h!]
  \centering
  \setlength{\tabcolsep}{2pt}
  \begin{tabular}{cc}
    \subfloat[]{\includegraphics[scale=0.45]{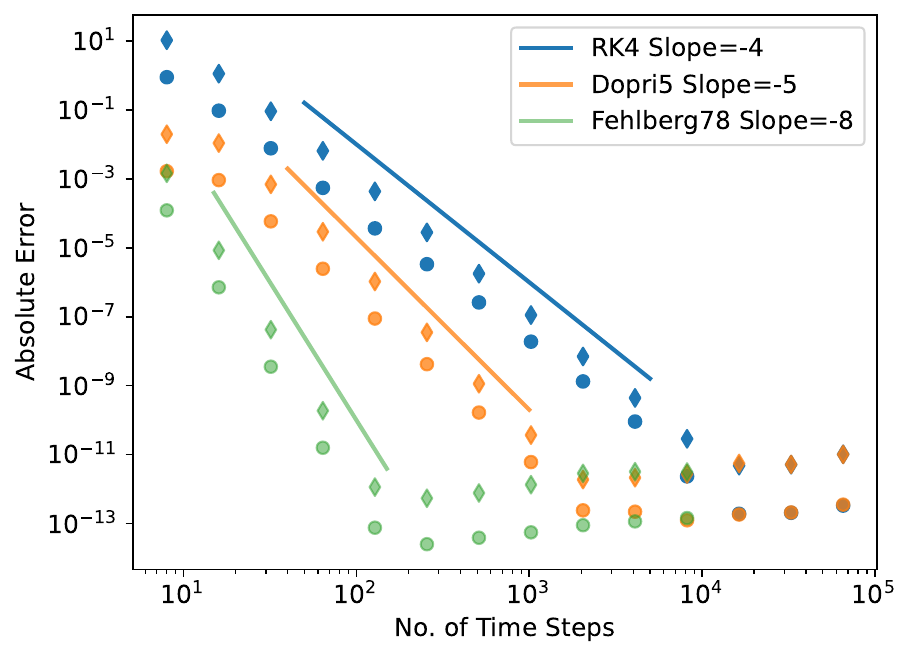}} &       \subfloat[]{\includegraphics[scale=0.45]{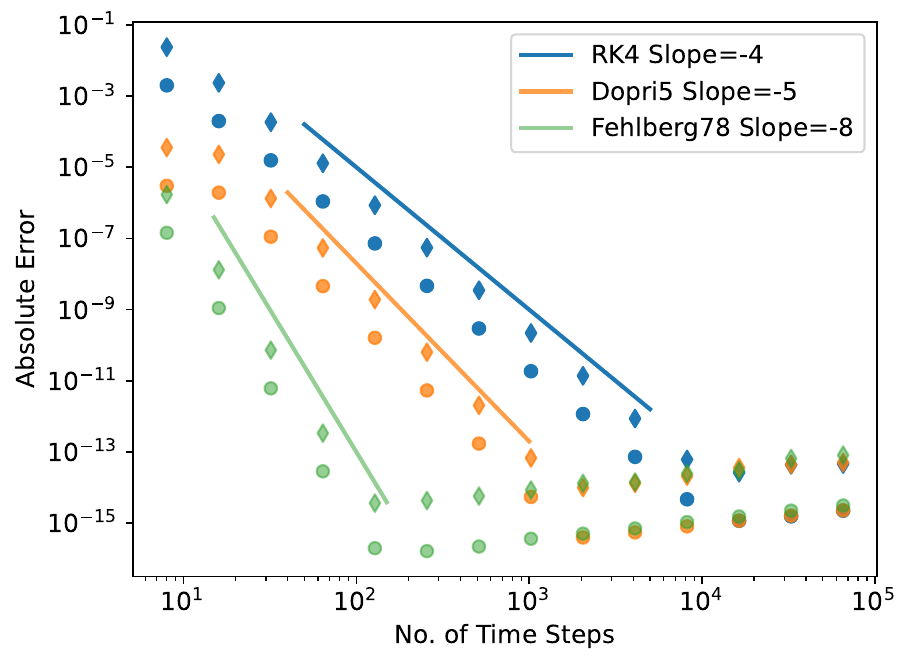}} \\
    \subfloat[]{\includegraphics[scale=0.45]{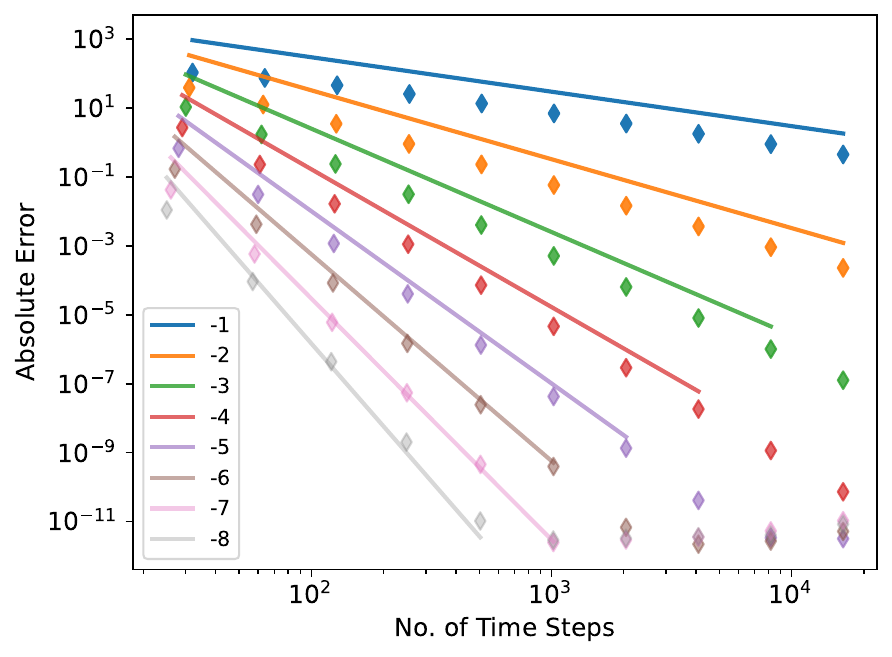}}   & \subfloat[]{\includegraphics[scale=0.45]{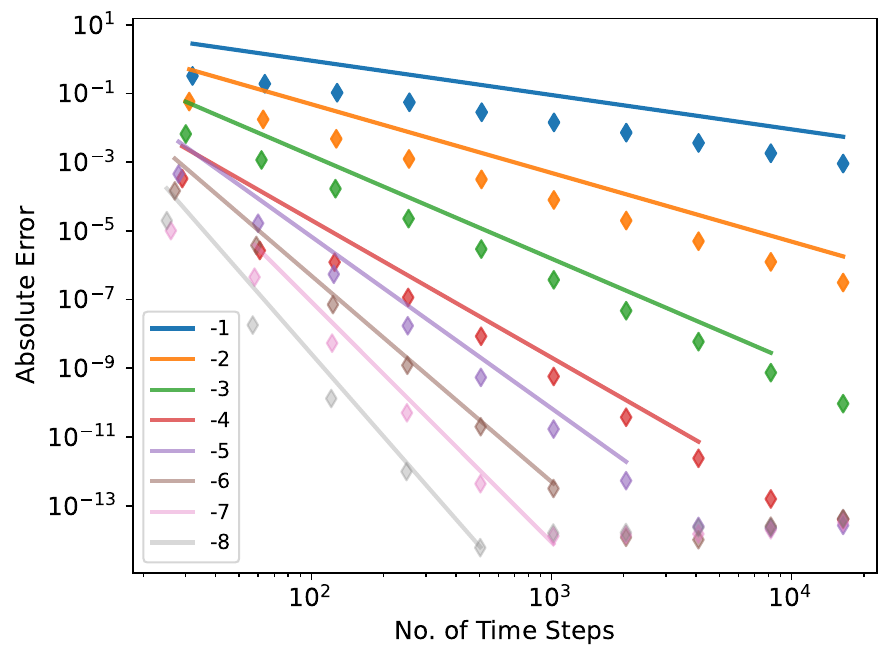}}
  \end{tabular}
  \caption{\textbf{Numerical convergence of time-integration schemes with increasing numbers of time steps.
    (a,b)} $L_\infty (\blacklozenge)$ and $L_2 (\bullet)$ error norms for the exponential dynamics of Eq.~(\ref{eq:exp}) (\textbf{a}) and the sigmoidal dynamics of Eq.~(\ref{eq:sig}) (\textbf{b}) with different one-step multi-stage methods (colors, inset legend). Solid lines indicate the theoretically expected slopes.
    \textbf{(c,d)} $L_\infty (\blacklozenge)$ error norms for the exponential
    (\textbf{c}) and sigmoidal (\textbf{d}) dynamics solved using Adams--Bashforth with different numbers of steps ($1\ldots 8$, colors, inset legend). Solid lines indicate the theoretically expected slopes.
  }
  \label{fig:cgs}
\end{figure}

We benchmark the implementation for correctness on problems with known analytical
solution. In order to generate many such problems, we consider the parametric
families of exponential and sigmoidal dynamics defined by:
\begin{subequations}
  \begin{align}
    \frac{\partial u(t)}{\partial t}&=xy\,\mathrm{e}^t, \label{eq:exp} \quad (x,y)\in [0,1]\times[0,1],\\
    \frac{\mathrm{d} u(t)}{\mathrm{d} t} &= \frac{1}{1+\mathrm{e}^{-t}}\left(1-\frac{1}{1+\mathrm{e}^{-t}}\right). \label{eq:sig}
  \end{align}
\end{subequations}
Hence, the exponential family contains as many different ODEs as the number of
discretization points in space. We use the Odeint--OpenFPM interface presented above
to solve this problem with the analytical solution as initial condition at $t_0=-5$.
We compare the computed solutions with the analytical solution at final time $t_f=5$
and compute the infinity norm $\|\mathbf{e}\|_\infty = \max(|e_1|,\ldots ,|e_n|)$ and
the Euclidean $L_2$ norm $\|\mathbf{e}\|_2 = \sqrt{e_1^2+\ldots +e_n^2}$ of the
absolute error $\mathbf{e}=\mathbf{u}-\mathbf{u}_{\mathrm{exact}}$ across all $n$
time steps. Methods with fixed time-step size are tested for different $\delta t$ to
verify their order of convergence.

We first test the fixed-step multi-stage methods Runge--Kutta 4, Runge--Kutta--Dopri
5, and Runge--Kutta--Fehlberg 78 (see Table \ref{Tab:steppers}) for the exponential
dynamics in Eq.~(\ref{eq:exp}). Figure~\ref{fig:cgs}a shows the convergence plots for
increasing numbers of time steps. All methods converge with the expected order, as
indicated by the solid lines, until machine precision is reached for the
\texttt{double} data type used, and finite-precision
round-off errors start accumulating. The same is observed in Fig.~\ref{fig:cgs}b for the sigmoidal dynamics of Eq.~(\ref{eq:sig}).

Next, we test the fixed-step Adams--Bashforth implementation as an example of a multi-step method. We solve the same problems using 1 to 8 steps, resulting in convergence orders between 1 and 8. This is verified for the exponential and sigmoidal dynamics in Figs.~\ref{fig:cgs}c and \ref{fig:cgs}d, respectively. In all cases, the theoretically optimal order of convergence is observed down to machine precision.

All tests are repeated for different numbers of OpenFPM processes (1, 2, 6, 24) to confirm that the results are the same, regardless of the degree of parallelism. Taken together, this confirms the correctness of the implementation of our Odeint  \emph{state-type} and \emph{Algebra} for distributed OpenFPM data types.

\subsection*{Scalability Benchmarks}\label{sec:usecase2}

We test the parallel scalability of the present software for multi-stage methods,
which incur additional communication overhead in a distributed-memory parallel
program when the ODEs result from spatial discretization. We again test the
fixed-step time integrators Runge--Kutta 4, --Dopri 5, and --Fehlberg 78, and
additionally --Dopri 5 with adaptive time steps. As a baseline, we use a native
implementation of Runge--Kutta 4 in OpenFPM without using Odeint and the associated \emph{Algebra}.

We solve the exponential dynamics given in Eq.~(\ref{eq:exp}) on $512\times 512$ points that discretize the domain $(x,y)\in [0,1]^2$ with equal spacing.
This amounts to 262,144 ODEs from the exponential family, which are
distributed among the processes using OpenFPM's domain decomposition and solved in parallel. We perform a strong scaling measurement with a constant problem size distributed across increasing numbers of processes (1 process per CPU core). The measurements are taken on a cluster of Intel Xeon E5-2680v3 CPUs @2.5\,GHz with each node containing 24 (2$\times$12) cores with shared memory. The nodes in the cluster are connected using a 4-lane FDR InfiniBand network (14\,Gb/s per lane) with a latency of \SI{0.7}{\micro\second} for message passing using the OpenMPI library.

\begin{figure}[h!]
  \setlength{\tabcolsep}{2pt}
  \begin{tabular}{cc}
    \subfloat[]{\includegraphics[scale=0.45]{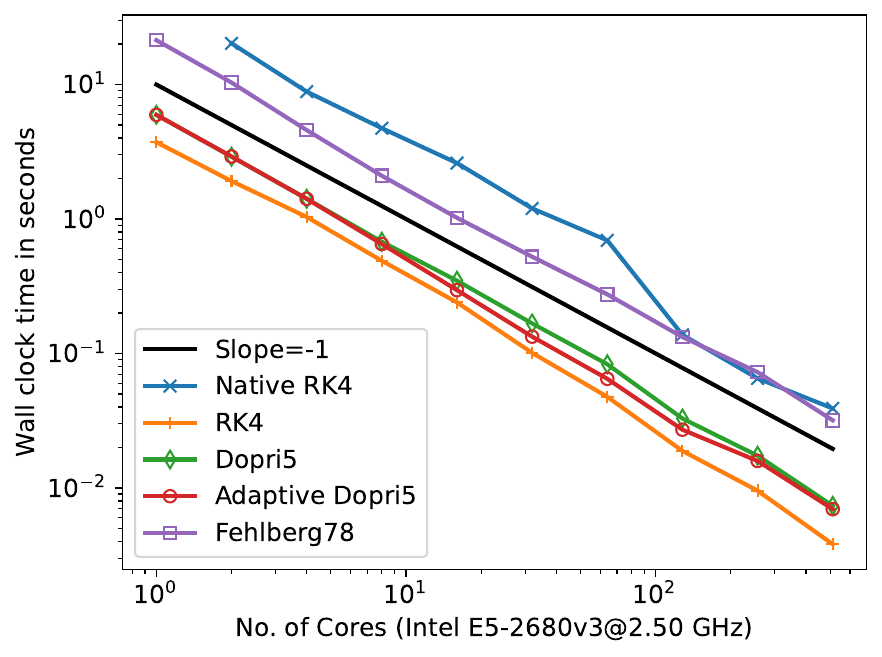}} &
    \subfloat[]{\includegraphics[scale=0.45]{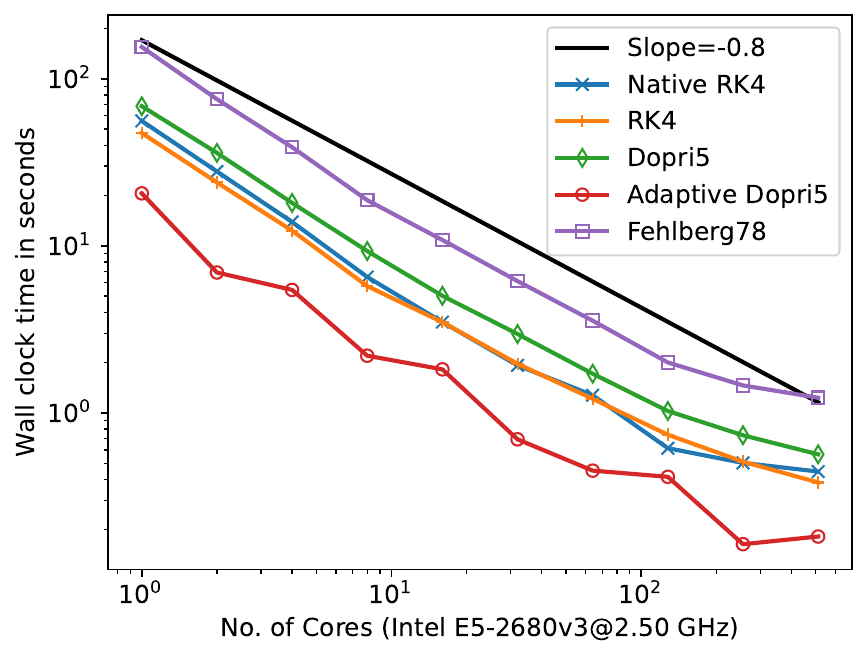}} \\
    \subfloat[]{\includegraphics[scale=0.37]{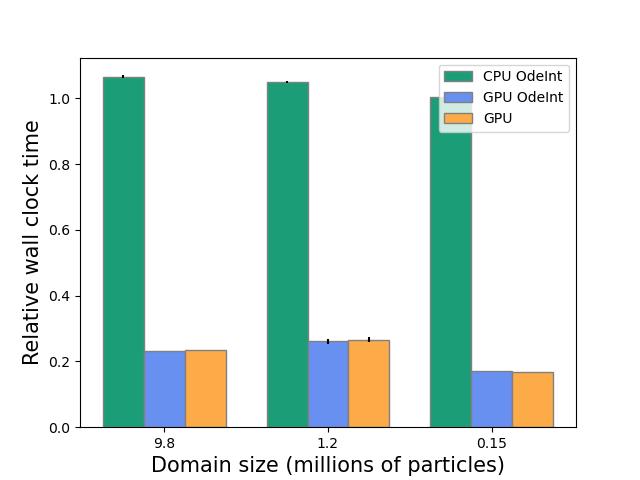}} &
    \subfloat[]{\includegraphics[scale=0.25]{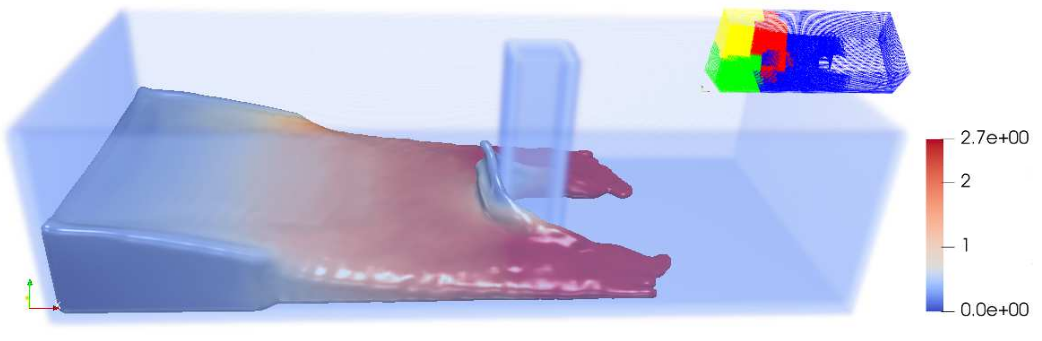}\vspace{5mm}}
  \end{tabular}
  \caption{\textbf{Strong scaling of the OpenFPM+Odeint time integration schemes
      with increasing numbers of CPU cores and on a GPU.
    (a)} Average wall-clock times in seconds over three independent runs of solving the exponential dynamics from Eq.~(\ref{eq:exp}) using different one-step multi-stage methods (colors, inset legend, error bars below symbol size).
    The solid black line indicates the optimal speed-up.
    \textbf{(b)} Average wall-clock times in seconds over 3 independent runs for
    the 3D Gray-Scott Eq.~(\ref{eq:gs}) using different one-step multi-stage
    methods (colors, inset legend, error bars below symbol size). The solid black line indicates the speed-up for 80\% parallel efficiency.
    \textbf{(c)} Average wall-clock speedups for the SPH dam break case, normalized to the runtimes reported for the reference CPU implementation without Odeint from \citet{INCARDONA2019155} (=1.0). All results are averaged over three independent runs (error bars show standard deviation) for different numbers of particles (bar groups). Speedups are given for a single Nvidia GeForce RTX 4090 GPU (blue bars), compared with running the code without the present Odeint interface on the same GPU (orange bars), and running and OpenFPM+Odeint code on 32 cores of an AMD Ryzen Threadripper 3990X CPU (green bars).
  \textbf{(d)}  Visualization of the fluid, colored by velocity magnitude (color bar), of the SPH dam break case with 1.2 million particles at time 0.43\,s. The OpenFPM domain decomposition onto four processes is shown in the inset figure (one color per process subdomain).}
  \label{fig:sc}
\end{figure}

The results are shown in Fig.~\ref{fig:sc}a. The native OpenFPM implementation is
almost an order of magnitude slower than the Odeint+OpenFPM implementation. This is
likely due to the performance gains from optimizing the on-the-fly computation of
stages in Odeint~\citep{odeint2015ahnert}. Since there is no spatial coupling in this
problem, we find close to ideal (solid black line) scaling of the computational time for all tested steppers. This establishes the ideal baseline scalability in the absence of spatial coupling.

To test how the performance changes when spatial coupling and corresponding communication overhead is introduced, we consider the 3D Gray-Scott reaction-diffusion problem that couples the ODEs in space:
\begin{equation}
  \frac{\partial }{\partial t}
  \begin{bmatrix}
    C_0(t,x,y,z)\\C_1(t,x,y,z)
  \end{bmatrix}  =
  \begin{bmatrix}
    d_1\mathrm{\Delta}_{(x,y,z)}C_0-C_0 C_1^2+F(1-C_0)\\
    d_2\mathrm{\Delta}_{(x,y,z)}C_1+C_0 C_1^1-(F+K)C_0
  \end{bmatrix}
  \label{eq:gs}
\end{equation}
for the parameters $d_1=2\cdot 10^{-4}$, $d_2=10^{-4}$, $F=0.053$, and $K=0.014$. The Laplace operators in space $\mathrm{\Delta}_{(x,y,z)}$ are discretized using the DC-PSE method~\citep{schrader2010discretization} as implemented in the PDE template expression system of OpenFPM~\citep{singh_c_2021}.
We use DC-PSE to solve the Gray-Scott model in the 3D cube $[0,2.5]^3$ with periodic boundary conditions on a regular Cartesian grid of $64\times64\times64$ points, time stepping with $\delta t=1$ until final time $t_f=20$.
The wall-clock times for different multi-stage schemes are plotted in Fig.~\ref{fig:sc}b.
We find a parallel efficiency of about 80\% up to 512 CPU cores (solid black line) for all tested
multi-stage schemes. Inter-processor communication of stages and of spatial
differential operators are performed inside the \emph{System} functor using OpenFPM's MPI
backend. Here, the native OpenFPM implementation has similar performance as the
OpenFPM+Odeint implementation, suggesting the bottleneck to be the spatial
derivatives computation.
Adaptive time stepping does not incur a significant
scalability penalty, while raw wall-clock times roughly halve. \rev{Taken together, these results show that the present OpenFPM--Odeint interface layer does not introduce a performance bottleneck, while providing OpenFPM with access to more elaborate, adaptive time steppers.}

While involving non-trivial time stepping, the previous test case was still relatively simple in terms of the equations that are solved. We therefore next quantify the performance of the preent software in a more complex, real-world test case. For this, we consider the classic dam-break scenario simulated using Smoothed Particle Hydrodynamics (SPH). This simulates the weakly compressible fluid mechanics of water slushing around a U-profile beam in a rectangular tank. This involves non-trivial geometries and free-surface flows. In this simulation, the water body is discretized with irregularly spaced Lagrangian particles $p$ whose velocities $\bm{v}_p$ and densities $\rho_p$ evolve according to:
\begin{subequations}
  \begin{align}
    \frac{d\bm{v}_p}{dt} &= - \!\!\sum_{q \in \mathcal{N}(p) } m_q \left(\frac{P_p + P_q}{\rho_p \rho_q} + \Pi_{pq} \right) \nabla W(\bm{x}_q - \bm{x}_p) + \bm{g}\, , \label{eq:sph1} \\
    \frac{d\rho_p}{dt} &=  \sum_{q \in \mathcal{N}(p) } m_q \bm{v}_{pq} \cdot \nabla W(\bm{x}_q - \bm{x}_p)\, , \label{eq:sph2} \\
    P_p &= \frac{1}{\gamma}c_\text{sound}^2 \|\bm{g}\|_2 h_\text{swl} \rho_{0} \left[ \left( \frac{\rho_p}{\rho_{0}} \right)^{\!\!\gamma} - 1 \right] \label{eq:statesph}  \, .
  \end{align}
\end{subequations}
Here, $h_\text{swl}$ is the initial height of the fluid, $\gamma=7$, and $c_\text{sound}=20$~\citep{Monaghan:1992}.
$\mathcal{N}(p)$ is the set of all particles within a cutoff radius of $2 \sqrt{3}h$ from
$p$, where $h$ is the distance between nearest neighbors.
$W(\bm{x})$ is the classic cubic SPH kernel~\cite{Monaghan:1992}, and $\bm{g}$ is the gravitational acceleration. The relative velocity between particles $p$ and $q$ is \mbox{$\bm{v}_{pq} = \bm{v}_p - \bm{v}_q$},
$\nabla W(\bm{x}_q - \bm{x}_p)$ is the analytical gradient of the kernel $W$ centered at particle $p$ and evaluated at the location of particle $q$.
The equation of state (Eq.~\ref{eq:statesph}) relates the hydrostatic pressure $P_p$ with the density $\rho_p$, where $\rho_0$ is the density of the fluid at \mbox{$P=0$}. The viscosity term $\Pi_{pq}$ is defined as:
\begin{equation}
  \Pi_{pq} =
  \begin{cases} - \frac {\alpha \bar{c}_{pq} \mu_{pq} }{\bar{\rho}_{pq} } & \bm{v}_{pq} \cdot \bm{r}_{pq} > 0\, ,
    \\ 0 & \bm{v}_{pq} \cdot \bm{r}_{pq} < 0\, .
  \end{cases}
\end{equation}
The vector $\bm{r}_{pq} = \bm{r}_p - \bm{r}_q$ points from particle $q$ to particle $p$, and the constants are defined as: $ \mu_{pq} = \frac{h \bm{v}_{pq} \cdot \bm{r}_{pq}}{\|\bm{r}_{pq}\|_2^2 + \eta^2} $, with fluid viscosity $\eta$, and $\bar{c}_{pq} = c_\text{sound}\sqrt{\|\bm{g}\|_2 h_{\text{swl}}}$.
We solve these equations using the SPH method implemented in OpenFPM until a final simulated time of 0.5\,s using velocity-Verlet time stepping with one explicit Euler step every 40 steps and step size computed for a CFL number of 0.2. As a baseline, we compare with the hand-written OpenFPM CPU implementation from \citet{INCARDONA2019155}.
We first quantify how much overhead the present Odeint interface adds over the manually implemented time integration used in the reference code (= speedup 1.0). The results in Fig.~\ref{fig:sc}c show that the overhead is below 7\% (between 0.3\% and 6.6\%, depending on problem size; green bars) for particle numbers ranging from 150,000 to 9.8 million. The measurements were done on 32 cores of an AMD Ryzen 3990X CPU. A visualization of the solution at time 0.43\,s is shown in Fig.~\ref{fig:sc}d, along with the automatically determined domain decomposition onto four processes (one per CPU core). This shows the applicability of the present OpenFPM--Odeint integration to a realistically complex application case on the CPU.

After having established the baseline overhead introduced by the present OpenFPM--Odeint interface on the CPU, we quantify the speedup achieved on a single Nvidia GeForce RTX 4090 consumer GPU. The present software natively supports parallel computing on GPUs through the GPU \emph{state-type}s as described above. Again, we compare with a hand-written reference implementation (orange bars in Fig.~\ref{fig:sc}c) and the implementation using the present Odeint interface (blue bars) for time stepping. for all problem sizes tested, the code is about 5 times faster on the GPU than on 32 CPU cores. The higher GPU speedup for the smallest problem size is likely due to lower data-transfer overhead.
The Odeint interface does not add any significant overhead on the GPU when compared with the hand-written GPU code (orange bars). Importantly, the GPU version of the present code did not require writing any CUDA by hand. All GPU targeting is done automatically by the OpenFPM backend as previously described~\citep{Incardona:2023a}.

\section*{(2) Availability}
\vspace{0.5cm}

We designed and implemented an interface between Boost Odeint~\citep{odeint2015ahnert} and OpenFPM \citep{INCARDONA2019155}. This
interface enables compact codes for scalable time integration on parallel computers
with performance portability between CPU and GPU clusters. The presented software
implementation is based on a custom state type and an internally distributed algebra
for Odeint. These objects were implemented using the templates and distributed data
types from the OpenFPM library. The interface is bi-directional, making distributed
OpenFPM types and network communication available in Odeint objects, and making the
Odeint time integrators available in OpenFPM's domain-specific language for
differential equations \citep{singh_c_2021}. This endows the template
expression language of OpenFPM with native primitives for time derivatives, which can then be evaluated using Odeint's wealth of methods.

The presented software rests on custom Odeint \emph{state-type} classes for the
distributed OpenFPM data structures, complete with their transparently distributed
Odeint \emph{Algebra}s, including CUDA and HIP GPU versions.
We derived this design from analyzing the architecture of Odeint, and it provides
distributed \emph{state-type}s for OpenFPM distributed vectors of different dimensionality.
The presented template algebra makes the OpenFPM-distributed \emph{state-type}s compatible with the generic stepper algebra of Odeint. As a result, all explicit time-stepping schemes of Odeint become available in OpenFPM, including adaptive, high-order, and multi-step schemes. They are available for all OpenFPM data containers with time derivatives encapsulated in a \emph{System} functor that also transparently handles inter-process communication as needed and discretizes spatial differential operators via OpenFPM template expressions \citep{singh_c_2021}.
This makes coding new OpenFPM-based solvers more efficient and less error-prone.

We tested the present implementation for correctness and scalability. The results
verified the correct convergence orders of all schemes, up to the finite machine
precision. We demonstrated that our implementation scales ideally in the absence of
inter-process communication and has a parallel efficiency of about 80\% (strong
scaling, 512 CPU cores) with the typical communication overhead of a coupled PDE
problem in 3D solved using multi-stage schemes. The additional Odeint abstractions did not add detectable overhead over a native OpenFPM implementation and in some cases were even faster, due to the highly optimized code of Odeint.

In summary, the present software implementation simplifies the use of time-stepping methods in scientific problem-solving environments, such as the OpenPME environment
\citep{OpenPME1,OpenPME2}, on distributed CPU and GPU computers. This has the potential of extending autotuning systems for spatial discretization methods \citep{khouzami_jocs21} to the time dimension, as changing the time stepper, or the step size, amounts to a single codeline change, which can be implemented in an autotuning framework.
The presented system allows for rapid rewriting of distributed and parallel numerical solvers with minimal changes to the source code, while maintaining scalability and performance portability.

\section*{Operating systems}

x86\_64/AMD64: Linux, macOS \& Unix-like, Window only with Cygwin (partial support); ARM64: macOS (min.~macOS 13 Ventura).

\section*{Programming languages}
C++14, CUDA (min.~9.2, for Nvidia GPU support), HIP (min.~5.0.0, for AMD GPU support)

\section*{Minimal hardware requirements}
x86\_64 or AMD64 CPU for Linux/Unix/Windows systems; ARM64 CPU for macOS. Minimum resources in all cases: 1 core, 1\,GB RAM, 15\,GB disk space, text terminal access. Optionally: a supported GPU from Nvidia or AMD.

\section*{Software dependencies}
See Table~\ref{Tab:dependencies}.

\begin{table}[!ht]
  \centering
  \captionof{table}{Software dependencies of OpenFPM\label{Tab:dependencies}}
  \begin{tabular}{|l|p{4cm}|l|l|}
    \hline
    \textbf{Dependency} & \textbf{Used by} & \textbf{Optional} & \textbf{Version} \\ \hline\hline
    Open MPI & openfpm\_vcluster & NO & 4.1.6 \\ \hline
    METIS & openfpm\_pdata & YES (or ParMETIS) & 5.1.0 \\ \hline
    ParMETIS & openfpm\_pdata, openfpm\_numerics & YES (or METIS) & 4.0.3 \\ \hline
    BOOST & openfpm\_data, openfpm\_vcluster, openfpm\_io, openfpm\_pdata, openfpm\_numerics & NO & 1.84.0 \\ \hline
    zlib & openfpm\_io & NO & 1.3.1 \\ \hline
    HDF5 & openfpm\_io & NO & 1.14.3 \\ \hline
    Vc & openfpm\_data & NO & 1.4.4 \\ \hline
    libhilbert & openfpm\_data & NO & master \\ \hline
    HIP & openfpm\_devices & Yes & ~ \\ \hline
    alpaka & openfpm\_devices & Yes & ~ \\ \hline
    OpenBLAS & openfpm\_numerics & NO & 0.3.26 \\ \hline
    suitesparse & openfpm\_numerics & NO & 5.7.2 \\ \hline
    Eigen & openfpm\_numerics & Yes (or Petsc) & 3.4.0 \\ \hline
    Blitz++ & openfpm\_numerics & NO & 1.0.2 \\ \hline
    Algoim & openfpm\_numerics & NO & master \\ \hline
    PETSc & openfpm\_numerics & Yes (or Eigen) & 3.20.5 \\ \hline
  \end{tabular}
\end{table}

\section*{List of code contributors}

Code contributors in alphabetical order are:
\begin{itemize}
  \item Pietro Incardona
  \item Landfried Kraatz
  \item Abhinav Singh
  \item Serhii Yaskovets
\end{itemize}

\section*{Software location:}

{\bf Code repository}

\begin{description}[noitemsep,topsep=0pt]
  \item[Name:] GitHub
  \item[Persistent identifier:] \url{https://github.com/mosaic-group/openfpm}
  \item[Licence:] 3-clause BSD
  \item[Date published:] 14/03/2025
\end{description}

\section*{Documentation language}
English

\section*{(3) Reuse potential}

With the OpenFPM \emph{state-type}s and \emph{Algebra} described here, a user can
implement Odeint solvers for dynamical systems using OpenFPM's embedded DSL for
differential equations~\citep{singh_c_2021}. Any time integrator provided by Odeint
can be used to approximate time derivatives in OpenFPM code. The performance portability and scalability of the OpenFPM data structures are inherited.

The present implementation is currently
limited to the explicit steppers available in the Odeint library and to fields of
dimensionality $\leq$6. While Odeint's
abstractions allow for defining arbitrary custom steppers, our algebra is currently
limited to steppers requiring no more than 15 inline methods. Further, implicit and stiff Odeint time integrators are not currently supported, as they are tied to uBLAS, which does not support generic types or parallelism. Also, support for
SUNDIALS~\citep{gardner2022sundials} could be added in the future as another
time-integration backend.

\rev{Nevertheless, the presented implementation has already proven instrumental to solve active hydrodynamic equations in complex-shaped 3D geometries~\citep{Singh2023}. This has led to the discovery of novel wrinkling instabilities in biological active fluids~\citep{singh_prr2023,Alam2024}. The governing equations of active hydrodynamics are challenging to solve, requiring extensive numerical experimentation and frequent changes to the code. This benefited from the rapid code rewriting afforded by the portable encapsulation of the present OpenFPM--Odeint interface.}

In order to implement an ODE solver, the right-hand side $\mathcal{F}$ needs to be
provided as an Odeint \emph{System} object with the signature of the \emph{state}
class. The \emph{System} function or functor is called by the time stepper at each
stage. It is parameterized by the templated \emph{state-type} and the data type of
its time derivative. Although it is unlikely that the solution state and its time
derivative have different data types, Odeint allows for them to be specified
separately in order to provide the most general implementation possible. Finally, the
current time $t$ is a parameter to the \emph{System}. Using these parameters, the
\emph{System} evaluates $\mathcal{F}(t,\mathbf{u}(t))$ for the current state \rev{$\mathbf{u}_t$} at time $t$.

\begin{lstlisting}[caption={Odeint \emph{System} functor for the 3D Gray-Scott problem in OpenFPM.},label={lst:system-functor},language=C++]
  //Templated type of the time derivative operator
  template<typename Laplacian_type>
  struct System_Functor
  {
    Laplacian_type &Lap
    double K = 0.053, F = 0.014, d1 = 2e-4, d2 = 1e-4; //Physical contants

    System_Functor(Laplacian_type &Lap) : Lap(Lap)
    {}

    void operator()(const state_type_2d_ofp &u, state_type_2d_ofp &dudt,
    const double t ) const
    {
      //Get OpenFPM distributed domain
      ofp_dist_vector_type &Domain= *(ofp_dist_vector_type*) DistVecPointer;

      auto C=getV<Conc>(Domain) ; //Get the property

      //Get the data from the current state.
      C[0]=u.data.get<0>();
      C[1]=u.data.get<1>();

      //Compute the time derivate using PDE expressions.
      Domain.ghost_get<Conc>();
      dudt.data.get<0>() = d1*Lap(C[0]) - C[0] * C[1] * C[1] + F - F * C[0];
      dudt.data.get<1>() = d2*Lap(C[1]) + C[0] * C[1] * C[1] - (F+K) * C[1];
    }
  };
\end{lstlisting}

Listing \ref{lst:system-functor} exemplifies the usage of the \emph{System} functor for
the 3D Gray-Scott reaction-diffusion problem considered in the scalability benchmarks above.
This example also illustrates that it is straightforward to integrate templated
OpenFPM operators for spatial derivatives into the ODE system (Laplacian
\texttt{Lap}, Lines 5, 8, 25, 26). The \emph{System} functor can also contain any
additional code that is to be run for a right-hand-side evaluation, for example for
performance profiling. Since the data as well as the operations are transparently
distributed by OpenFPM in a multi-node or multi-GPU computer, boundary layers
(``ghost layers'') may need to be communicated between processes. This MPI
communication is performed in the OpenFPM {\texttt{Domain}} class. A reference to the
OpenFPM {\texttt{Domain}} object is therefore required in Line 15. Data from the
\emph{state} is then stored in the respective property of the OpenFPM
{\texttt{Domain}} compatible with the spatial operators computed. Required
inter-process communication of stage data is transparently performed during right-hand-side evaluation from within the \emph{System} functor (Line 24), guaranteeing consistent \emph{state}s to Odeint at all times.

After defining the \emph{System}, any Odeint time stepper can be used for time integration in the {\texttt{main()}} program, as shown in Listing \ref{lst:time-stepper-example} for the \emph{System} from Listing \ref{lst:system-functor}.

\begin{lstlisting}[caption={An OpenFPM program using an Odeint time stepper.},label={lst:time-stepper-example},language=C++]
  //Initialize OpenFPM data strucures and spatial derivatives
  auto C=getV<Conc>(Domain);
  Laplacian Lap(Domain);

  //Define stepper type
  odeint::runge_kutta4<state_type_2d_ofp, double, state_type_2d_ofp, double, odeint::vector_space_algebra_ofp> Odeint_rk4;

  //Declare and initialize state
  state_type_2d_ofp u;
  u.data.get<0>()=C[0];
  u.data.get<1>()=C[1];

  //Initialize System functor with spatial Laplacian R.H.S.
  System_Functor<Laplacian> System(Lap);

  //Invoke RK4 stepper for t=[0,20] with time step 0.1
  double t =0, tf = 20, dt = 0.1;
  size_t steps = integrate_const(rk4(), System, u, t, tf, dt);
\end{lstlisting}

The stepper used in Line 6 is one of the built-in steppers of Odeint (see Table
\ref{Tab:steppers}). Before starting time integration, we initialize the \emph{state}
in Lines 9--11 using the 2D OpenFPM CPU vector state type
{\texttt{state\_type\_2d\_ofp}}. The two components of the vector are then
set to the initial condition. The correct distributed \emph{Algebra} for this state
type is instantiated in Line 6 from {\texttt{odeint::vector\_space\_algebra\_ofp}} as
described above. We then initialize the system functor from
Listing~\ref{lst:system-functor} in Line 14. Finally, the time stepper is invoked to
run until a given final time in Line 18. Alternatively, the method {\texttt{do\_step()}} of the stepper can be called to perform a single time step only.

Adaptive time stepping with error-controlled steppers can be enabled by calling
{\texttt{integrate\_adaptive()}} with the desired tolerance. The stepper returns the
number time steps performed, and the state is updated in-place. Thanks to the use of
OpenFPM, the program in Listing \ref{lst:time-stepper-example} compiles and runs on
shared- and distributed-memory CPU systems, as well as on Nvidia and AMD GPUs and mutli-GPU clusters.

\section*{Funding statement}

This work was funded in parts by the German Research Foundation (DFG, Deutsche
Forschungsgemeinschaft) under grants SB-350008342 (``OpenPME'') (author P.I.) and
EXC-2068, Cluster of Excellence ``Physics of Life'' (author I.F.S.), and by the
Federal Ministry of Education and Research of Germany (Bundesministerium f\"{u}r Bildung und Forschung, BMBF) under funding codes 031L0160 (project ``SPlaT-DM -- computer simulation platform for topology-driven morphogenesis'') (author A.S.) and the Center for Scalable Data Analytics and Artificial Intelligence (ScaDS.AI) Dresden/Leipzig (author A.S.).

\section*{Competing interests}

The authors declare that they have no competing interests.


\bibliographystyle{agsm}
\bibliography{bibliography}

\vspace{2cm}

\rule{\textwidth}{1pt}

{ \bf Copyright Notice} \\
Authors who publish with this journal agree to the following terms: \\

Authors retain copyright and grant the journal right of first publication with the work simultaneously licensed under a  \href{http://creativecommons.org/licenses/by/3.0/}{Creative Commons Attribution License} that allows others to share the work with an acknowledgement of the work's authorship and initial publication in this journal. \\

Authors are able to enter into separate, additional contractual arrangements for the non-exclusive distribution of the journal's published version of the work (e.g., post it to an institutional repository or publish it in a book), with an acknowledgement of its initial publication in this journal. \\

By submitting this paper you agree to the terms of this Copyright Notice, which will apply to this submission if and when it is published by this journal.

\end{document}